\documentclass{article}
\usepackage{dcase2019,amsmath,graphicx,url,times,booktabs, tabularx}
\usepackage{indentfirst}
\usepackage{siunitx}
\usepackage[all]{nowidow}

\title{Sound source detection, localization and classification using consecutive ensemble of CRNN models}

\name{S\l{}awomir Kapka\sthanks{Corresponding author.}, Mateusz Lewandowski}
\address{Samsung R\&D Institute Poland\\
      Artificial Intelligence\\
      Warsaw, 00-844, Poland \\
      \{s.kapka, m.lewandows4\}@samsung.com}

\begin{document}

\ninept
\maketitle

\begin{sloppy}

\begin{abstract}
\indent \indent In this paper, we describe our method for DCASE2019 task~3: Sound Event Localization and Detection (SELD). We use four CRNN SELDnet-like single output models which run in a consecutive manner to recover all possible information of occurring events. We decompose the SELD task into estimating number of active sources, estimating direction of arrival of a single source, estimating direction of arrival of the second source where the direction of the first one is known and a multi-label classification task. We use custom consecutive ensemble to predict events' onset, offset, direction of arrival and class. The proposed approach is evaluated on the TAU Spatial Sound Events 2019 - Ambisonic and it is compared with other participants' submissions.
\end{abstract}

\begin{keywords}
DCASE 2019, Sound Event Localization and Detection, CRNN, Ambisonics
\end{keywords}

\section{Introduction}
\label{sec:intro}
Sound Event Localization and Detection (SELD) is a complex task which naturally appears when one wants to develop a system that possesses spatial awareness of the surrounding world using multi-channel audio signals. This year, the task 3 from the IEEE AASP Challenge on Detection and Classification of Acoustic Scenes and Events (DCASE 2019) \cite{dcase2019web} concerned the SELD problem. SELDnet introduced in \cite{Adavanne2018_JSTSP} is a single system of a good quality designed for the SELD task, and the slight modification of SELDnet was set as the baseline system \cite{Adavanne2019_DCASE} during the DCASE 2019 Challenge. Solely based on \cite{Adavanne2018_JSTSP} and \cite{Adavanne2019_DCASE}, we develop a novel system designed for the task 3 from the DCASE 2019 Challenge.

In our work, we follow the philosophy that if a complex problem can be split into simpler ones, one should do so. Thus we decompose the SELD task with up to 2 active sound sources into the following subtasks:

\begin{itemize}
  \item estimating the number of active sources (\emph{noas}),
  \item estimating the direction of arrival of a sound event when there is one active sound source (\emph{doa1}),
  \item estimating the direction of arrival of a sound event when there are two active sound sources and we posses the knowledge of the direction of arrival of one of these sound events, which we will call an \emph{associated event} (\emph{doa2}),
  \item multi-label classification of sound events (\emph{class}).
\end{itemize}
For each of this subtasks, we develop a SELDnet-like convolutional recurrent neural network (CRNN) with a single output. We discuss it in detail in section \ref{sec:architect}. Given such models, we develop a custom consecutive ensemble of these models. This allows us to predict the events' onset, offset, direction of arrival and class, which we discuss in detail in section \ref{sec:ensemble}. Due to the sequential nature of generating predictions in our system, errors in models' predictions may cascade, and thus an overall error may cumulate. Despite this drawback, our system acquire very good results on the TAU Spatial Sound Events 2019 - Ambisonic database. We discuss the results in detail in section \ref{sec:results}.

\begin{figure}[t]
  \centering
  \centerline{\includegraphics[width=\columnwidth]{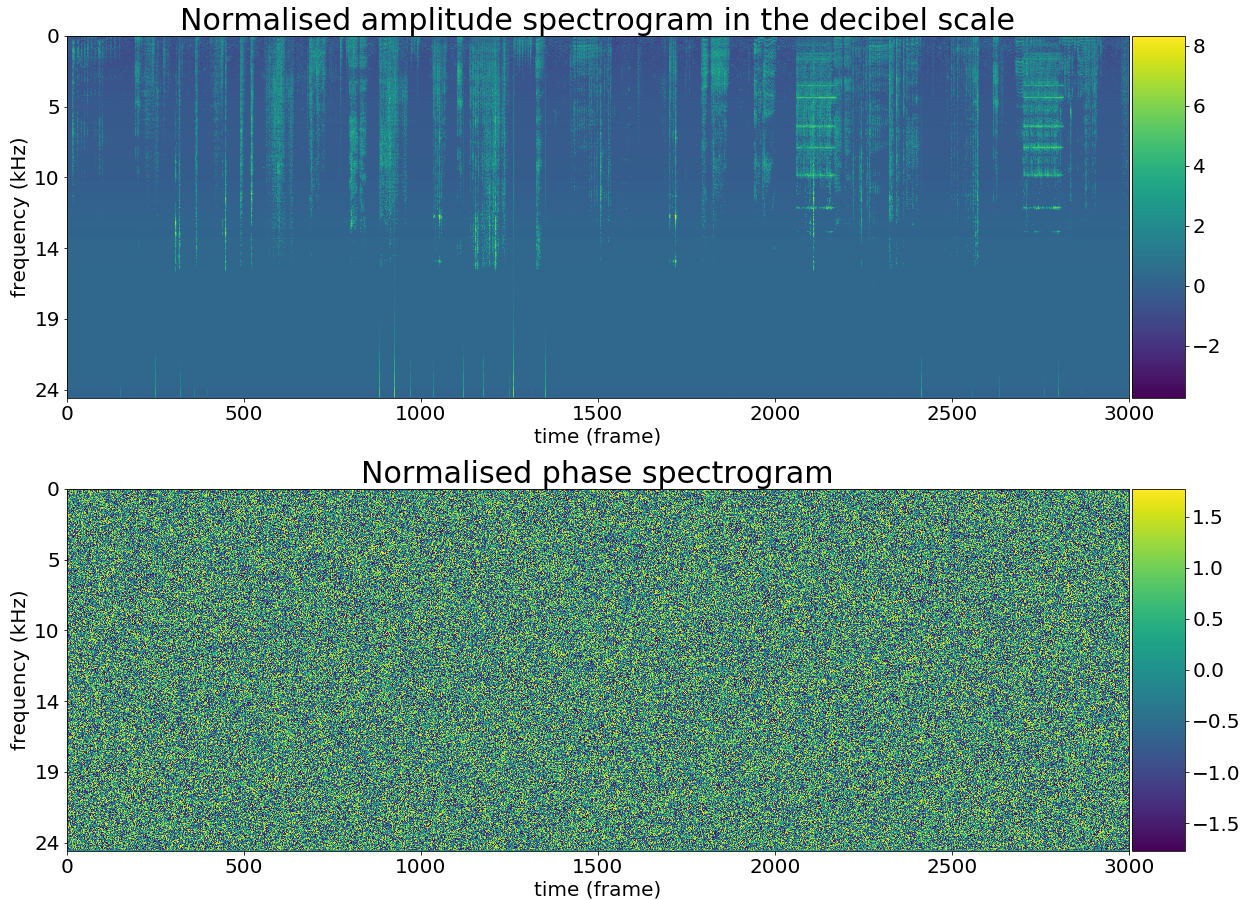}}
  \caption{An example of the normalised amplitude spectrogram in the decibel scale and the normalised phase spectrogram obtained from the first \emph{foa} channel from some randomly selected recording. The horizontal and vertical axes denote frame numbers and frequencies respectively obtained from the STFT. Note that the values from the legends on the right are dimensionless due to the normalization used in the preprocessing.}
  \label{fig:spectrograms}
\end{figure}

\section{Features}
\label{sec:features}
The DCASE 2019 task 3 provides two formats of the TAU Spatial Sound Events 2019 dataset: first order ambisonic (\emph{foa}) and 4 channels from a microphone array (\emph{mic}) \cite{Adavanne2019_DCASE}.  In our method we only use the ambisonic format.

Each recording is approximately 1 minute long with sampling rate of 48k. We use the short time Fourier transform (STFT) with Hann window. We use the window of length 0.4s and hop of length 0.2s in STFT to transform a raw audio associated to each \emph{foa} channel into the complex spectrogram of size 3000x1024. If audio is longer than 1 minute, we truncate spectrograms. If an audio is shorter than 1 minute, we pad them with zeros.

From each complex spectrogram we extract its module and phase point-wise, that is amplitude and phase spectrograms, respectively. We transform amplitude spectrograms to the decibel scale. Finally, we standardize all spectrograms frequency-wise to zero mean and unit variance, to obtain spectrograms as in Figure \ref{fig:spectrograms}.

In summary, from each recording we acquire 4 standardized amplitude spectrograms in the decibel scale and 4 standardized phase spectrograms corresponding to 4 \emph{foa} channels.

\section{Architecture}
\label{sec:architect}

As mentioned in the introduction, each of the subtasks (\emph{noas}, \emph{doa1}, \emph{doa2} and \emph{class}) has its own SELDnet-like CRNN. Each of these models is a copy of a single SELDnet node with just minor adjustments so that it fits to the specific subtask and for the regularization purpose.

Each of these models takes as an input a fixed length subsequence of decibel scale amplitude spectrograms (in case of \emph{noas} and \emph{class} subtasks) or both decibel scale amplitude and phase spectrograms (in case of \emph{doa1} and \emph{doa2} subtasks) from all 4 channels.

In each case, the input layers are followed by 3 convolutional layer blocks. Each block is made of a convolutional layer, batch norm, relu activation, maxpool and dropout. The output from the last convolutional block is reshaped so that it forms a multivariate sequence of a fixed length. In the case of \emph{doa2}, we additionaly concatenate directions of arrivals of associated events with this multivariate sequence. Next, there are two recurrent layers (GRU or LSTM) with 128 units each with dropout and recurrent dropout. Next layer is a time distributed dense layer with dropout and with the number of units depending on subtask.

Lastly, depending on a subtask, the model has a different output. For \emph{noas}, the model has just a single time distributed output that corresponds to the number of active sources (0, 1 or 2). For \emph{doa1} and \emph{doa2}, the models have 3 time distributed outputs that corresponds to cartesian xyz coordinates as in \cite{Adavanne2018_JSTSP}. Cartesian coordinates are advantageous over spherical coordinates in this task due to their continuity. Lastly, for \emph{class}, the model has 11 time distributed outputs corresponding to 11 possible classes. We present the detailed architecture in Table \ref{tab:parameteres}.

\begin{table*}[t]
  \caption{The architecture and the parameters of the networks}
  \label{tab:lcnn_architecture}
  \centering
  \begin{tabular}{*6l}
    \toprule
     \textbf{Layer Type} & \textbf{Parameters}  & \emph{noas}  & \emph{doa1} & \emph{doa2} & \emph{class}     \\
    \midrule
    Input & Shape & $256 \times 1024 \times 4$ & $128 \times 1024 \times 8$ & $128 \times 1024 \times 8$ & $128 \times 1024 \times 4$\\
    
    \textit{ConvBlock*} & Pool & 8 & 8 & 8 & 8\\
    \textit{ConvBlock*} & Pool & 8 & 8 & 8 & 8\\
    \textit{ConvBlock*} & Pool & 4 & 4 & 4 & 4\\
    
    Reshape & Sequence length $\times$ features & $256 \times -1$ & $128 \times -1$ & $128 \times -1$ & $128 \times -1$\\
    Doa2 input & Is used & False & False & True & False\\
    Concatenate & Is used & False & False & True & False\\
    \textit{RecBlock**} & Unit type & GRU & LSTM & GRU & GRU\\
    \textit{RecBlock**} & Unit type & GRU & LSTM & GRU & GRU\\
    TD Dense & Number of units & $16$ & $128$ & $128$ & $16$\\
    Dropout & Dropout rate & $0.2$ & $0.2$ & $0.2$ & $0.2$ \\
    TD Dense & Number of units & $1$ & $3$ & $3$ & $11$\\
    Activation & Function & linear & linear & linear & sigmoid\\

    \midrule
    \textit{*ConvBlock$(P)$}& \\
    \midrule
    Conv2D &  \multicolumn{5}{l}{$64$ filters, $3 \times 3$ kernel, $1\times 1$ stride, same padding} \\
    BatchNorm & --- \\
    Activation & \multicolumn{5}{l}{ReLu function} \\
    MaxPooling2D & \multicolumn{5}{l}{$1\times P$ pooling}\\
    Dropout & \multicolumn{5}{l}{$0.2$ dropout rate} \\
    \midrule
    \textit{**RecBlock$(U)$}& \\
    \midrule
    Recurrent & \multicolumn{5}{l}{$128$ recurrent units of type $U$, $0.2$ recurrent dropout rate} \\
    Activation & \multicolumn{5}{l}{$\tanh$ function} \\
    Dropout & \multicolumn{5}{l}{$0.2$ dropout rate} \\
    \bottomrule
  \end{tabular}
  \label{tab:parameteres}
\end{table*}

Depending on a subtask, we feed the network with the whole recordings or just their parts. For \emph{noas}, we feed all the data. For \emph{doa1}, we extract only those parts of the recordings where there is just one sound source active. For \emph{doa2}, we extract only those parts of the recordings where there are exactly two active sound sources. For \emph{class}, we extract those parts of the recordings where there are at least one active source.

As for the learning process, we used mean square error loss for the \emph{noas}, \emph{doa1}, \emph{doa2} subtasks and binary cross-entropy loss for the \emph{class} subtask. For all subtasks we initialised learning process using Adam optimizer with default parameters \cite{Adam}. The \emph{noas} and \emph{class} subtasks were learned for 500 epochs with exponential learning rate decay; every 5 epochs the learning rate were multiplied by 0.95. In \emph{doa1} and \emph{doa2} subtasks, we run learning process for 1000 epochs without changing the initial learning rate.

As for complexity, the \emph{noas}, \emph{doa1}, \emph{doa2} and \emph{class} have \num[group-separator={,}]{572129}, \num[group-separator={,}]{753603}, \num[group-separator={,}]{591555} and \num[group-separator={,}]{572299} parameters respectively, making total of \num[group-separator={,}]{2651634} parameters.

\section{Consecutive ensemble}
\label{sec:ensemble}
In this section, we introduce and describe the idea of the consecutive ensemble which is the core of our approach. This custom binding of our four models allows us to predict the events' onset, offset, direction of arrival and class.

\subsection{The algorithm}
\label{ssec:algorithm}

We assume that recordings have at most 2 active sound sources at once and the sound events occur on a 10 degrees resolution grid. In our setting, the audios after feature extraction have exactly 3000 vectors corresponding to the time dimension. Henceforth we will call these vectors as frames. The algorithm itself goes as follows:

1. We feed the features to the \emph{noas} network to predict the number of active sources (NOAS) in each frame.

2. We transform the predicted NOAS so that each recording starts and ends with no sound sources and the difference of NOAS between each frames is no greater than 1.

3. From the predicted NOAS we deduce the number of events, their onsets and the list of possible offsets for each event. If NOAS in two consecutive frames increases, then we predict that a new event happened at the second frame. If in two consecutive frames NOAS decreases, then we append the first frame to all events since last time NOAS was 0 as a possible offset.

4. In order to determine which offset corresponds to which event we use the \emph{doa1} network. We extract chunks (intervals of equal NOAS) of audio where the predicted NOAS equals 1 and we feed it to \emph{doa1} network. For each chunk where NOAS was 1 we predict the average azimuth and elevation, and we round it to the closest multiple of 10. If two consecutive chunks have the same azimuth and elevation then we conclude that the first event covered two chunks and the second event started and ended between those chunks. If two consecutive chunks have a different azimuth or elevation, then we conclude that the first event ended when the second chunk started and the second event continued in the second chunk.

5. To determine the remaining information about angles we need to predict the direction of arrival (DOA) of events that start and end while the associated event is happening. We feed the chunks where NOAS is 2 to the \emph{doa2} network with the second input being DOA of the associated event in cartesian xyz coordinates. Similarly as in step 4, we average the predicted results from chunks and round it to the closest multiple of 10.

6. Lastly, we predict the events' classes. If an event has chunks where the event is happening in an isolation (NOAS = 1), then all such chunks are feed to the \emph{class} network and the most probable class (using soft voting among frames) is taken as a predicted class. If an event has no such chunks, i.e. the event is only happening with an associated event, then such chunk (NOAS = 2) is fed to the network and two most probable classes are extracted. We choose the first one which does not equal to the class of the associated event.

\subsection{An example}
\label{ssec:example}

The algorithm itself may seem quite complex at first glance. Hence, we investigate here a concrete example. 

Given a recording constituting of 3000 vectors, we predict its NOAS in each frame as in Figure \ref{fig:noas}. For the sake of clarity we constrain only to a part of the recording. Consider a block with predicted NOAS as in the top plot from Figure \ref{fig:example2}. According to the step 3 from the algorithm, we predict that 3 events happened here: $E_1, E_2, E_3$ with 3 corresponding onsets $On_1, On_2, On_3$. Events $E_1$ and $E_2$ may end at $Off_1, Off_2$ or $Off_3$ and event $E_3$ may end at $Off_2$ or $Off_3$ (see the bottom plot from Figure \ref{fig:example2}). According to the step 4 from the algorithm, we predict DOA using \emph{doa1} in chunks from $On_1$ to $On_2$, from $Off_1$ to $On_3$ and from $Off_2$ to $Off_3$. Based on that we deduce the events' offsets as in Figure \ref{fig:example2}. Based on step 5 from the algorithm, we predict the DOA of chunk from $On_3$ to $Off_2$ using \emph{doa2} where the associated DOA is the DOA of $E_2$. Lastly we deduce classes of the events $E_1, E_2$ and $E_3$. According to the step 6 form the algorithm, we predict class of $E_1$ based on the chunk from $On_1$ to $On_2$, predict the class of $E_2$ based on chunks from $Off_1$ to $On_3$ and from $Off_2$ to $Off_3$. Finally, we predict the class of $E_3$ based on the chunk from $On_3$ to $Off_2$. If the predicted class of $E_3$ is the same as the class of $E_2$ then we predict it to be the second most probable class from the \emph{class} network.

\begin{figure}[t]
 \centering
 \centerline{\includegraphics[width=\columnwidth]{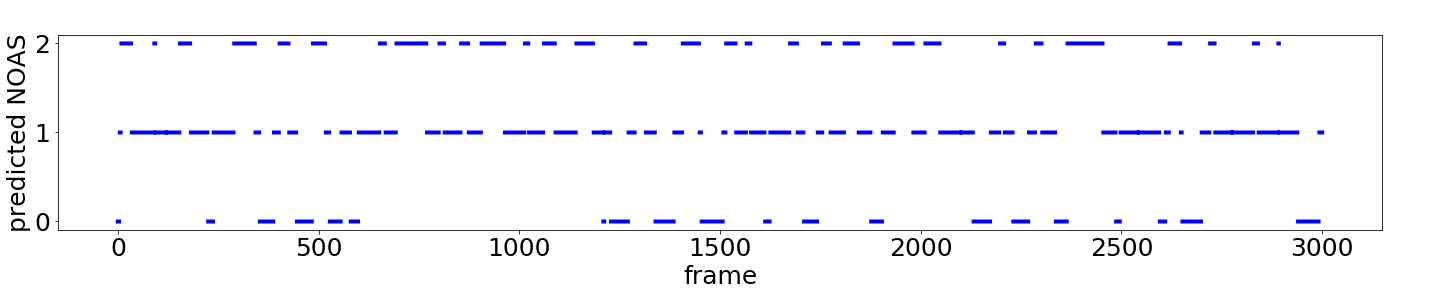}}
 \caption{The plot visualising the predicted number of active sources for some randomly selected recording.}
 \label{fig:noas}
\end{figure}

\begin{figure*}[t]
 \centering
 \centerline{\includegraphics[width=\textwidth]{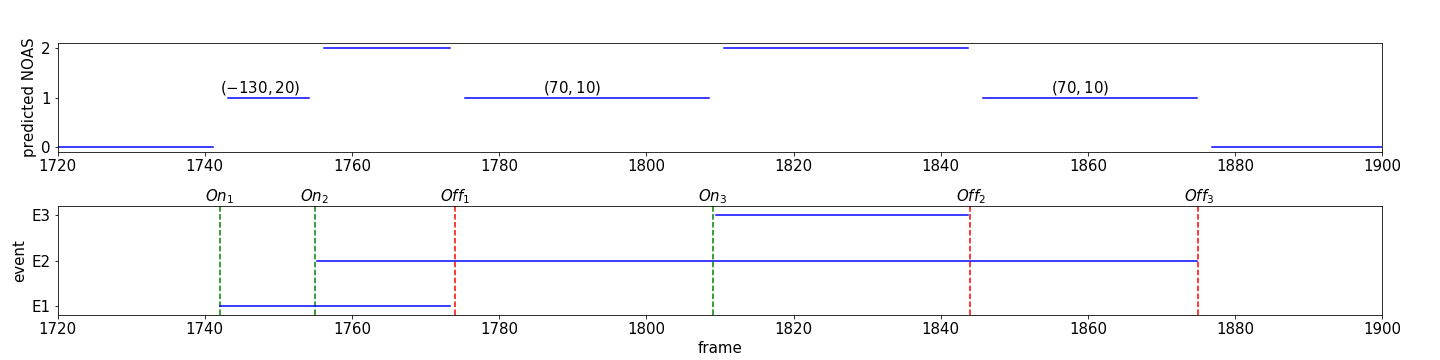}}
 \caption{Given the predicted NOAS from the part of some recording as in the top plot, we deduce that there are 3 events $E_1, E_2$ and $E_3$ with corresponding onsets denoted by the green lines in the bottom plot. Based on the predicted DOA, which we placed in the top plot above the segments, we deduce the events' offsets denoted by the red lines in the bottom plot.}
 \label{fig:example2}
\end{figure*}

\section{Results}
\label{sec:results}

We evaluate our results on TAU Spatial Sound Events 2019 - Ambisonic dataset. This dataset constitutes of two parts: the development and evaluation sets. The development part consists of 400 recordings with predefined 4-fold cross-validation and the evaluation part consists of 100 recordings. The results from this section relate to our submission \verb|Kapka_SRPOL_task3_2|.

\subsection{Development phase}
\label{ssec:develop}

As for the development part, we used 2 splits out of 4 for training for every fold using the suggested cross-validation even though validation splits do not influence the training process.

We show in Table \ref{tab:results} the averaged metrics from all folds for our setting and metrics for the baseline \cite{Adavanne2019_DCASE}. In order to demonstrate the variance among folds, we present in Table \ref{tab:test_splits} the detailed results on the test splits from each fold. The development set provides the distinction for the files where there is up to 1 active sound source at once (ov1) and where there are up to 2 (ov2). In Table \ref{tab:ov1_ov2} we compare metrics for the ov1 and ov2 subsets.

\begin{table}
\caption{The average results from all 4 splits.}
\scriptsize
\centering
\begin{tabular}{l*5c} 
 \cline{2-6}
 \multicolumn{1}{c}{} & Error rate & F-score & DOA error & Frame recall & Seld score \\
 \hline
 Train & 0.03 & 0.98 & 2.71 & 0.98 & 0.02 \\
 Val. & 0.15 & 0.89 & 4.81 & 0.95 & 0.08 \\ 
 Test & 0.14 & 0.90 & 4.75 & 0.95 & 0.08 \\ 
 \hline 
 Baseline & 0.34 & 0.80 & 28.5 & 0.85 & 0.22 \\ 
 \hline
\end{tabular}
\label{tab:results}
\end{table}

\begin{table}
\caption{The results on the test splits from each fold.}
\scriptsize
\centering
\begin{tabular}{l*5c} 
 \cline{2-6}
 \multicolumn{1}{c}{} & Error rate & F-score & DOA error & Frame recall & Seld score \\
 \hline
 Split 1 & 0.13 & 0.91 & 6.01 & 0.95 & 0.07 \\
 Split 2 & 0.16 & 0.88 & 6.01 & 0.95 & 0.09 \\ 
 Split 3 & 0.11 & 0.93 & 4.93 & 0.96 & 0.06 \\ 
 Split 4 & 0.17 & 0.86 & 5.89 & 0.96 & 0.10 \\ 
 \hline
\end{tabular}
\label{tab:test_splits}
\end{table}

\begin{table}
\caption{The results on the ov1 and ov2 subsets.}
\scriptsize
\centering
\begin{tabular}{l*5c} 
 \cline{2-6}
 \multicolumn{1}{c}{} & Error rate & F-score & DOA error & Frame recall & Seld score \\
 \hline
 ov1 & 0.07 & 0.94 & 1.28 & 0.99 & 0.04 \\
 ov2 & 0.18 & 0.87 & 7.96 & 0.93 & 0.11 \\ 
 \hline
\end{tabular}
\label{tab:ov1_ov2}
\end{table}

\subsection{Official results}
\label{ssec:official}

For the evaluation part, we used all 4 splits for training from the development set. We compare our final results with the selected submissions in Table \ref{tab:comparison}.

The idea of decomposing the SELD task into simpler ones proved to be a very popular idea among contestants. The recent two-stage approach to SELD introduced in \cite{Cao_oryginal} was used and developed further by many. The best submission using two-step approach \verb|Cao_Surrey_task3_4| \cite{Cao} obtained results very similar to ours. \verb|He_THU_task3_2| \cite{He} and \verb|Chang_HYU_task3_3| \cite{Chang} outperform our submission in SED metrics and DOA error respectively. However, our approach based on estimating NOAS first allows us to outperform all contestants in frame recall.

\begin{table}
\caption{The comparison of the selected submissions.}
\tiny
\centering
\begin{tabular}{*2l*4c} 
\hline
\textbf{Rank} & \textbf{Submission name} & \textbf{Error rate} & \textbf{F-score} & \textbf{DOA error} & \textbf{Frame recall} \\
\hline
1 & \verb|Kapka_SRPOL_task3_2| & 0.08 & 94.7 & 3.7 &  \textbf{96.8} \\
4 & \verb|Cao_Surrey_task3_4| & 0.08 & 95.5 & 5.5 & 92.2 \\
6 & \verb|He_THU_task3_2| &  \textbf{0.06} &  \textbf{96.7} & 22.4 & 94.1 \\
19 & \verb|Chang_HYU_task3_3| & 0.14 & 91.9 &  \textbf{2.7} & 90.8 \\
48 & \verb|DCASE2019_FOA_baseline| & 0.28 & 85.4 & 24.6 & 85.7 \\
\hline
\end{tabular}
\label{tab:comparison}
\end{table}

\section{Submissions}
\label{sec:submission}

Overall, we created 4 submissions for the competition:
\begin{itemize}
 \item ConseqFOA (\verb|Kapka_SRPOL_task3_2|),
 \item ConseqFOA1 (\verb|Kapka_SRPOL_task3_3|),
 \item ConseqFOAb (\verb|Kapka_SRPOL_task3_4|),
 \item MLDcT32019 (\verb|Lewandowski_SRPOL_task3_1|).
\end{itemize}

The first three submissions use the approach described in the above sections. The only difference is that ConseqFOA is trained on all four splits from development dataset. ConseqFOA1 is trained on splits 2,3,4. ConseqFOAb is trained on all splits but the classifier in this version was trained using categorical cross-entropy instead of binary cross-entropy loss.

Our MLDcT32019 submission uses a different approach. It works in the same way as the original SELDnet architecture but with the following differences:
\begin{itemize}
  \item We implemented the Squeeze-and-Excitation block \cite{Hu_2018_CVPR} after the last convolutional block. We pass the output from the last convolutional block through two densely connected neural layers with respectively 1 and 4 neurons, we multiply it with the output of the last convolutional block and we pass it further to recurrent layers.
  \item We set all dropout rates to $0.2$.
  \item We used SpecAugment \cite{specAug} as an augmentation technique to double the training dataset.
  \item We replaced recurrent layer GRU units with LSTM units.
\end{itemize}

\section{Conclusion}
\label{sec:conclusion}

We conclude that decomposing the SELD problem into simpler tasks is instinctive and efficient. However, we are aware that our solution has some serious limitations and it fails when one wants to consider a more general setup. For example when there are more than 2 active sources at once or when the grid resolution is more refined. Thus, we claim that the pursuit for universal and efficient SELD solutions is still open.

\section{Acknowledgement}
\label{sec:acknowledgement}

We are most grateful to Zuzanna Kwiatkowska for spending her time on careful reading with a deep understanding the final draft of this paper.

\pagebreak

\bibliographystyle{IEEEtran}
\bibliography{refs}
\end{sloppy}
\end{document}